# Tevatron Run I SUGRA Results

C. Pagliarone
*INFN-Pisa, V. Livornese 1291, 56010 Pisa, Italy*
*e-mail: pagliarone@fnal.gov*

(On the behalf of the CDF and D∅ Collaborations)

---

## Abstract

We present the most recent results of searches for physics beyond the Standard Model using the CDF and D∅ detector in the contest of Supersymmetry (SUSY) with Supergravity (SUGRA) constraints. All results described correspond to analysis performed using the past 1992-1996 Fermilab Tevatron Run I data (roughly 110pb$^{-1}$ per each experiment). In particular we report on searches for stop decay in tau channel assuming ℜ-Parity violation; searches for ℜ-Parity Violating LSP decays in di-muon plus 4 jets channel; searches for resonant slepton production in ℜ-Parity Violating mSUGRA; searches for mSUGRA in single electron channel assuming ℜ-Parity Conservation and searches for stop decay in 3-4 bodies.

---

## 1. Introduction

Although, at present, the Standard Model (SM) provides a remarkably successful description of known phenomena, there are plenty of aspects that we do not understand yet and that may suggest the SM to be most likely a low energy effective theory of spin-1/2 matter fermions interacting via spin-1 gauge bosons. An excellent candidate to a new theory, able to describe physics at arbitrarily high energies, is Supersymmetry (SUSY). SUSY is a large class of theoretical models based on the common assumption that there exist in nature a fermion-boson symmetry. In Supersymmetry fermions can couple to a sfermion and a fermion, violating lepton and/or baryon number. To avoid this problem, a discrete multiplicative quantum number, the ℜ-Parity, was introduced [1]: $\Re \equiv (-1)^{3B+L+2S}$. SUSY models can be constructed assuming either conservation (RPC) or violation (RPV) of this quantum number.

## 2. Search for ℜ-Parity Violating decay of stop search into taus

CDF searched for a pair produced scalar top squark decaying via non-zero ℜ-parity violating coupling $\lambda'_{333}$ to $\tilde{t}_1 \rightarrow \tau b$ [2]. The experimental signature of this process is two $\tau$ leptons and two b quarks in the final state. Events have been selected by requiring a lepton (e or μ) from $\tau \to \ell \nu_\ell \nu_\tau$, a hadronically decaying tau lepton and two jets. The principal background processes for this search are $Z \to \tau^+\tau^-$, W+jets, $t\bar{t}$, Drell-Yan and diboson events.

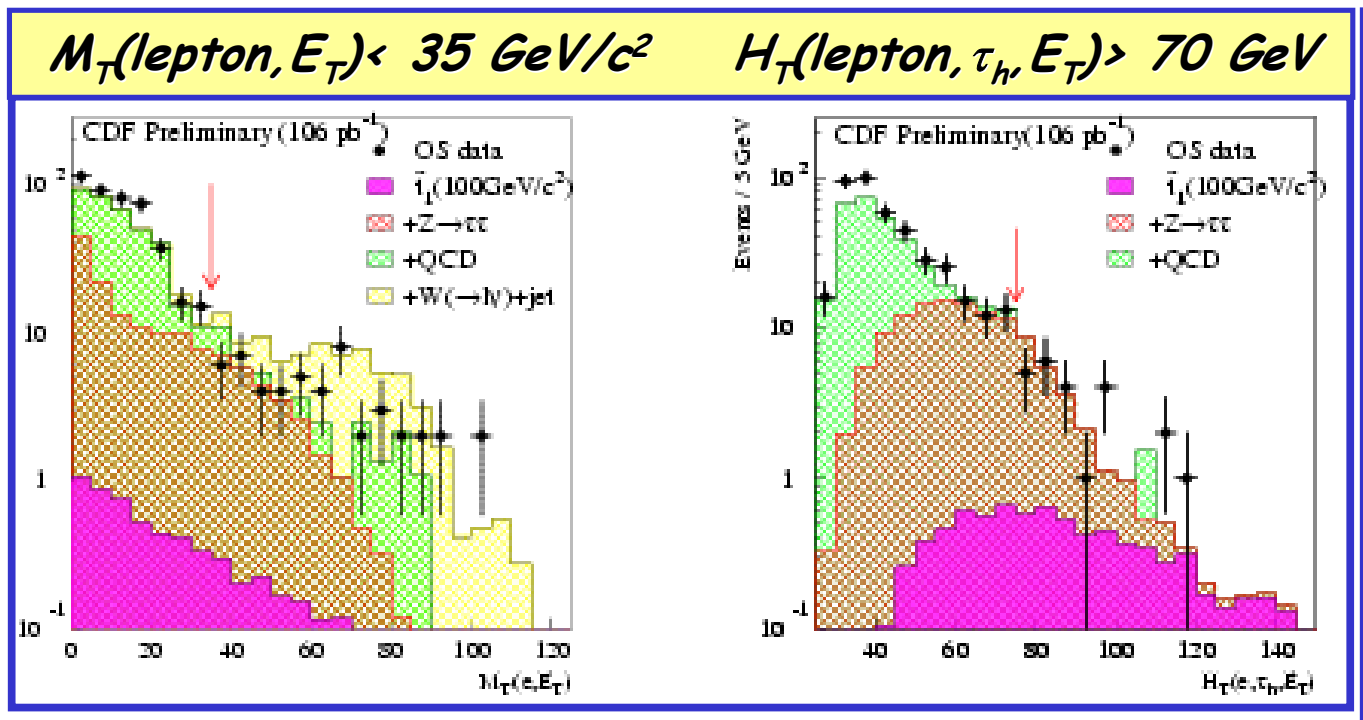

| channel | Background | Events |
|---|---|---|
| $e$ | 1.92 ± 0.18 | 0 |
| $\mu$ | 1.13 ± 0.13 | 0 |

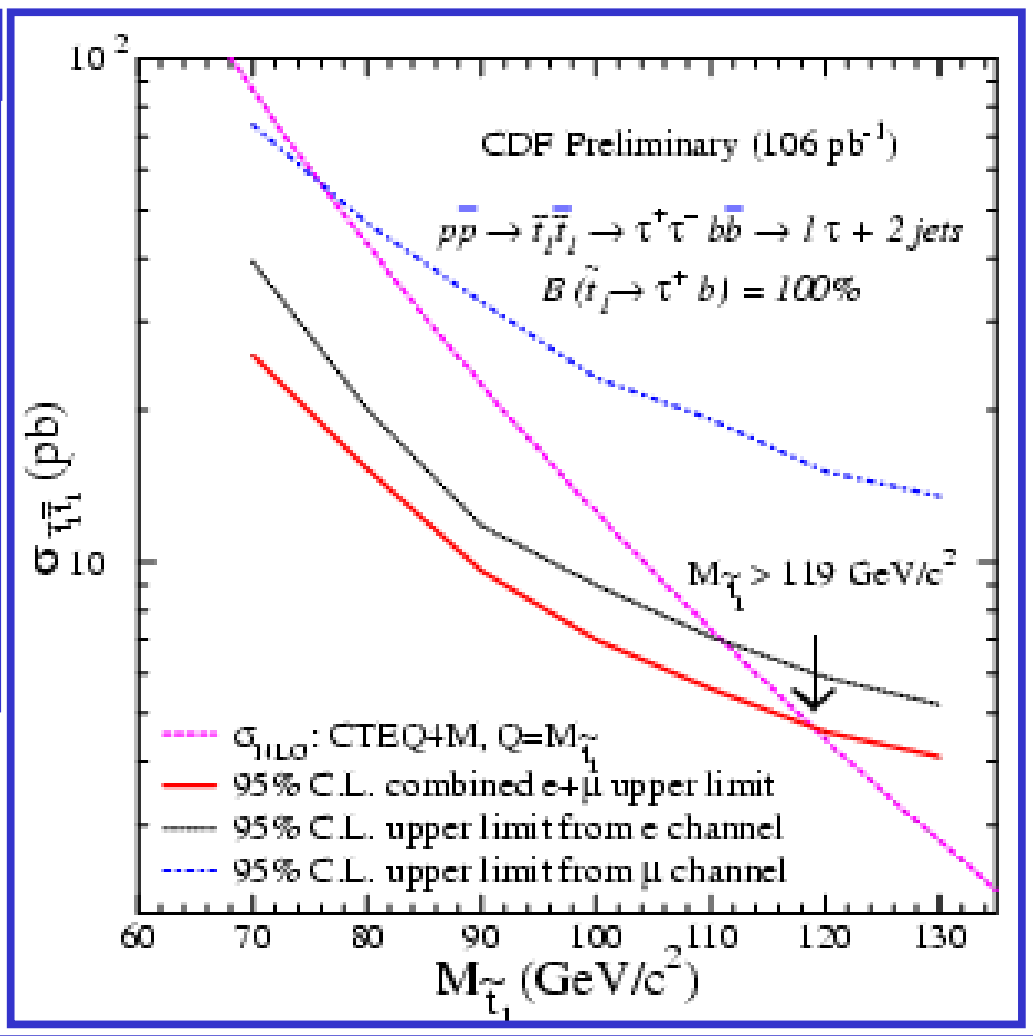

**FIG 1.** *Search for* ℜ*-Parity Violating decay of stop search into taus*

We observed, combining both the muon ($\tilde{t}_1\bar{\tilde{t}}_1 \to \tau^+\tau^- b\bar{b} \to \mu\tau_h b\bar{b} + X$) and the electron channel ($\tilde{t}_1\bar{\tilde{t}}_1 \to \tau^+\tau^- b\bar{b} \to e\tau_h b\bar{b} + X$), that no events passed the selection cuts.

This is consistent with the expected SM background of 1.92 ± 0.19 events in the electron channel and 1.13 ± 0.14 in the muon channel. A 95% *C.L.* lower limit on the stop quark mass have been set: M{$\tilde{t}_1$}> 119 GeV/c$^2$ (see figure 1) for a dominant $\lambda'_{333}$ coupling. The more recent and competitive result on the lower limit of the stop mass with this signature comes from the LEP experiment ALEPH [3].

## 3. Search for ℜ-Parity Violating LSP decays in di-muon and 4 jets

This analysis, performed from D∅ using an integrated luminosity of 77.5 $pb^{-1}$, complements a previous one searching for RPV decays of the LSP in the di-electron plus 4 jets channel [4], [5]. Each of the LSPs decays via lepton Flavor Violating (LFV) processes into a lepton and two jets: $\tilde{\chi}_1^0 \to e(\mu)qq'$. Events are selected by requiring the presence of at least four jets in the finals state ($E_T^{jet} > 15\ GeV$) and two muons having $P_T^{\mu(1)} > 15\ GeV$ and $P_T^{\mu(2)} > 10\ GeV$. The most important sources of background are $Z + jets$ and $t\bar{t}$ events. The number of background events expected, after all the cuts have been applied, is, for an integrated luminosity of 77.5 $pb^{-1}$, $0.18 \pm 0.04$. No events pass the cuts and a 95% C.L. limit can be set. In the context of mSUGRA with $A_0 = 0$, $\mu < 0$, $tg\beta = 2, 6$ the resulting bounds are expressed in the common gaugino versus scalar mass ($m_{1/2} \times m_0$) as shown in figure 2. In Figure 3 we compare the Run I limits, both for dimuon and dielectron channels, with the expected Run IIA exclusion limits assuming an integrated luminosity of $\approx 2\,fb^{-1}$.

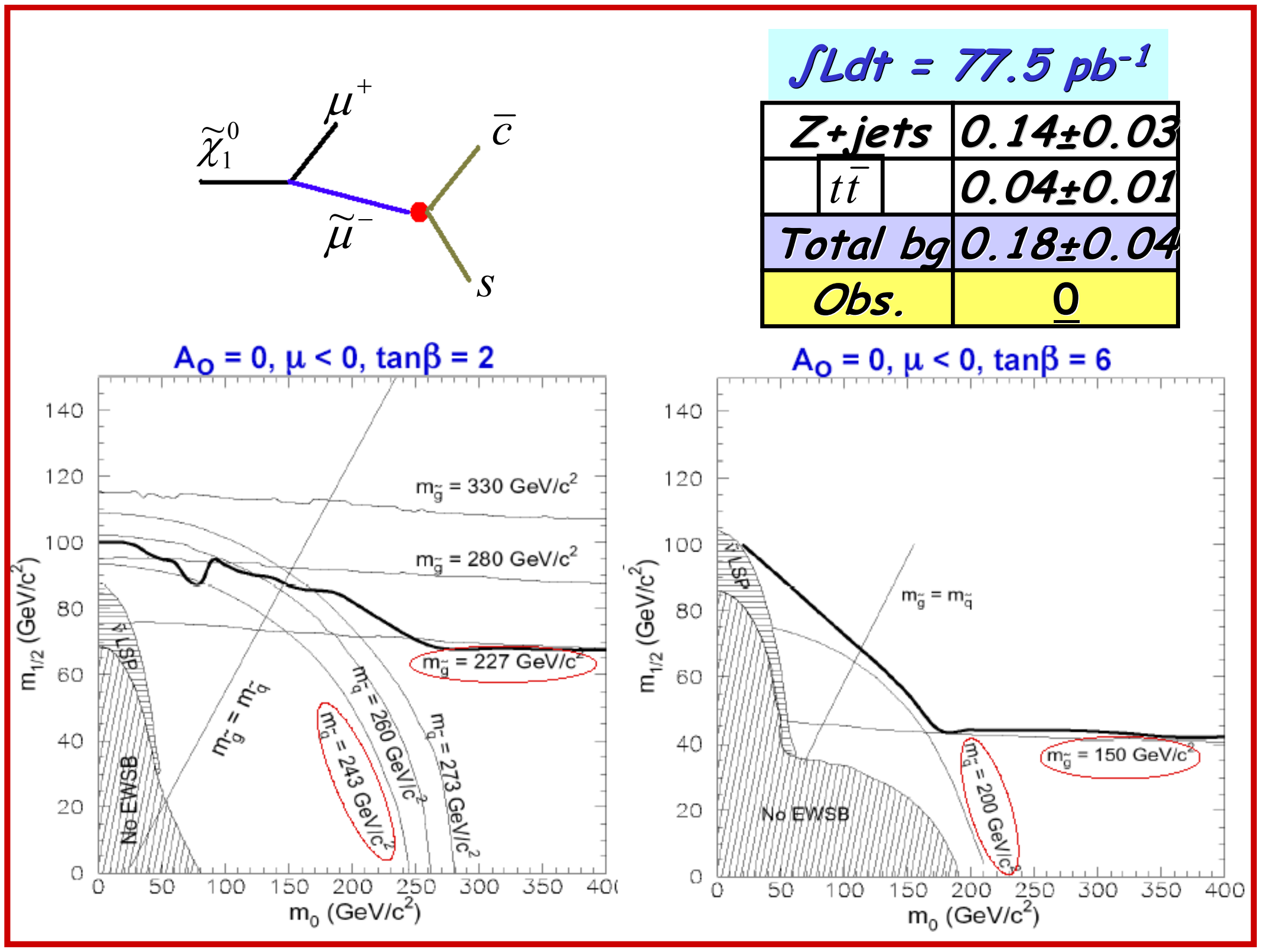

**FIG 2.** *Exclusion curves as function of* $m_{1/2} \times m_0$ *for* $tg\beta = 2$ *and* $tg\beta = 6$.

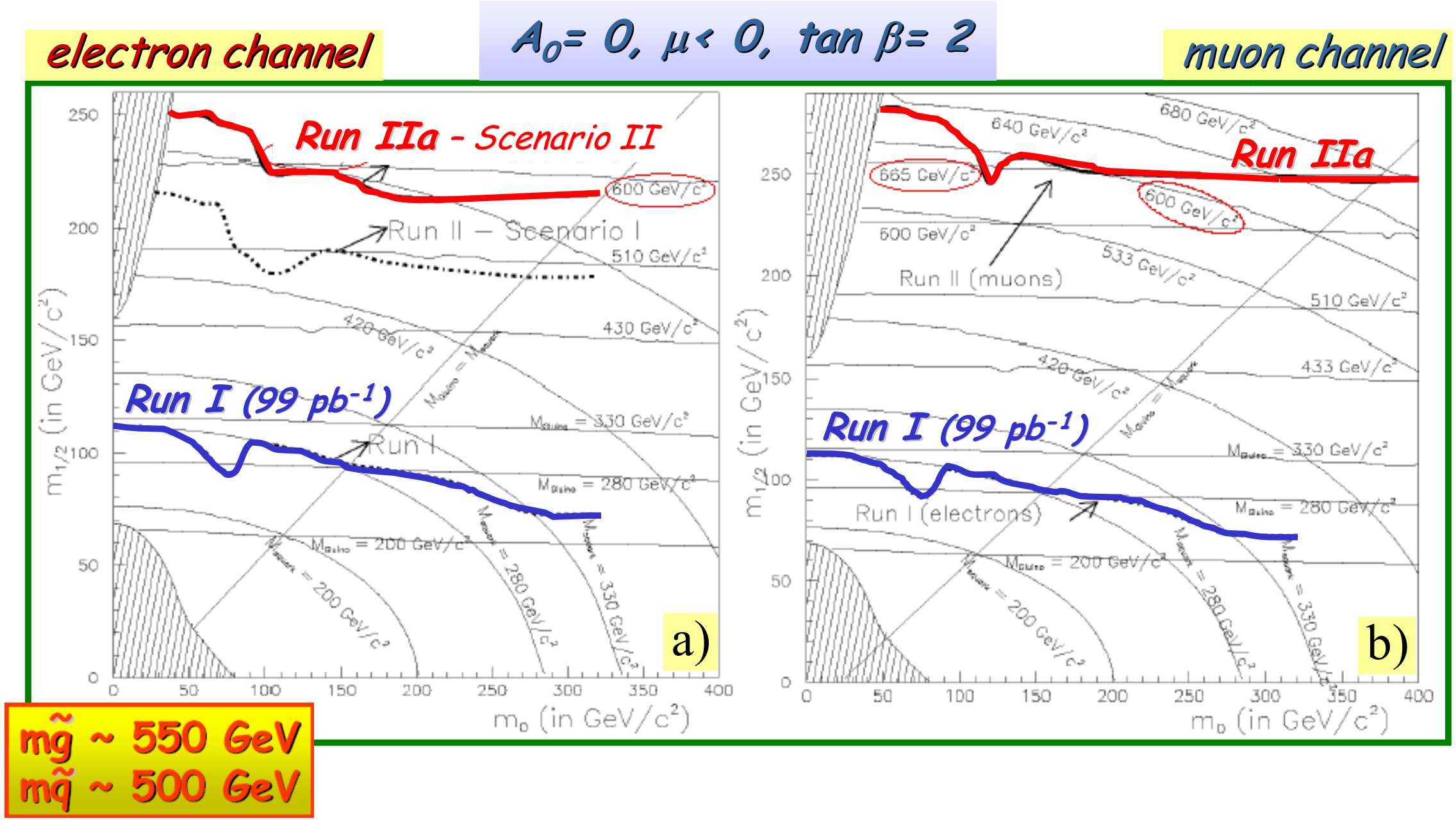

**FIG 3.** *Run II expected exclusion curves as function of* $m_{1/2} \times m_0$ *for* $tg\beta = 2$ *and* $tg\beta = 6$.

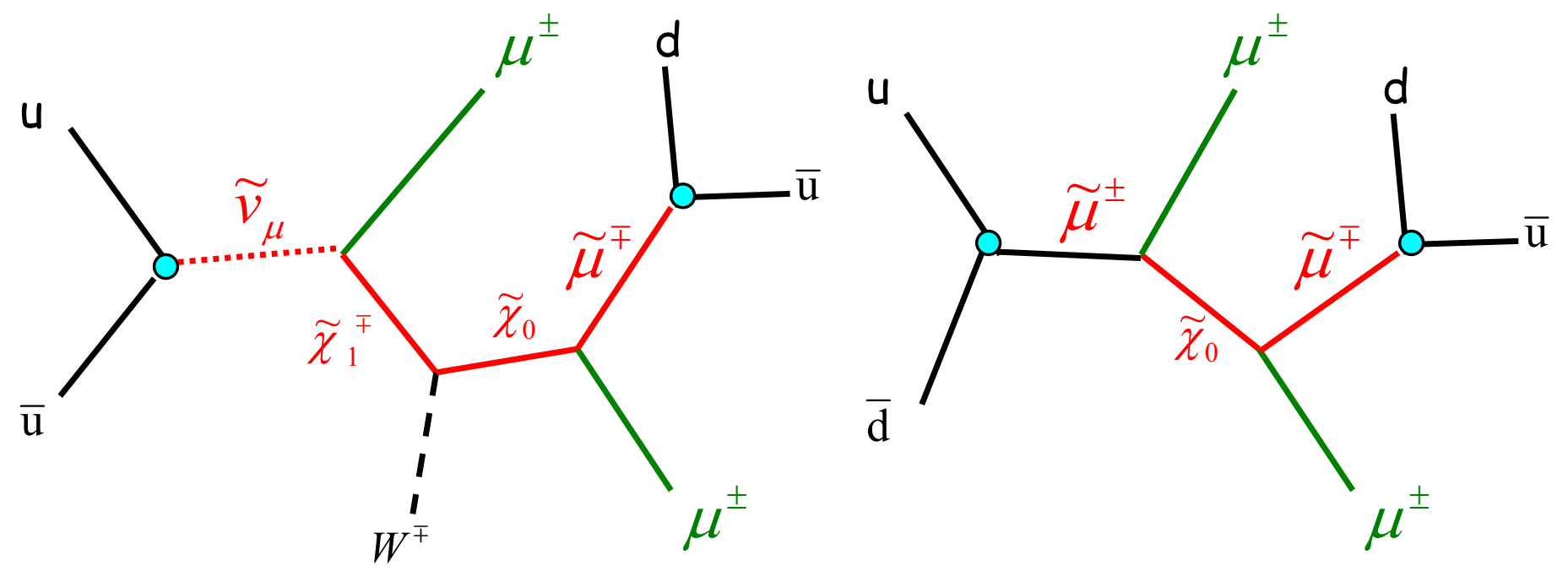

**FIG 4.** *Feynman diagrams for the resonant slepton production in mSUGRA models with R-Parity Violation with* $\lambda'_{211}$ *as a dominant coupling.*

## 4. Search for Resonant Slepton production in RPV mSUGRA

In this process the dominant coupling is $\lambda'_{211}$. This allows a resonant $\tilde{\mu}$ or $\tilde{\nu}_{\mu}$ production. The relative Feynman diagram are given in Figure 4. The signature for this process is then 2 muons plus two or more jets in the final state. The main background sources are represented by Drell-Yan processes, $t\bar{t}$ production, $Z+jets$ production and *WW+jets*. The analysis have been performed by D∅ using 94 $pb^{-1}$ of Run I data. Events are selected by requiring the presence of at least two jets with $E_T^{jet} > 20\ GeV$ and two muons with $P_T^{\mu} > 20\ GeV$. After all the analysis cuts have been applied 5 events are observed. The number of expected background events is: $5.34 \pm 0.07$. The resulting bound is once again expressed as function of gaugino versus scalar masses ($m_{1/2} \times m_0$). The limits are shown in figure 5 for the following sets of mSUGRA parameters: $A_0 = 0,\ \ \mu < 0,\ \ tg\beta = 2,\ 6$.

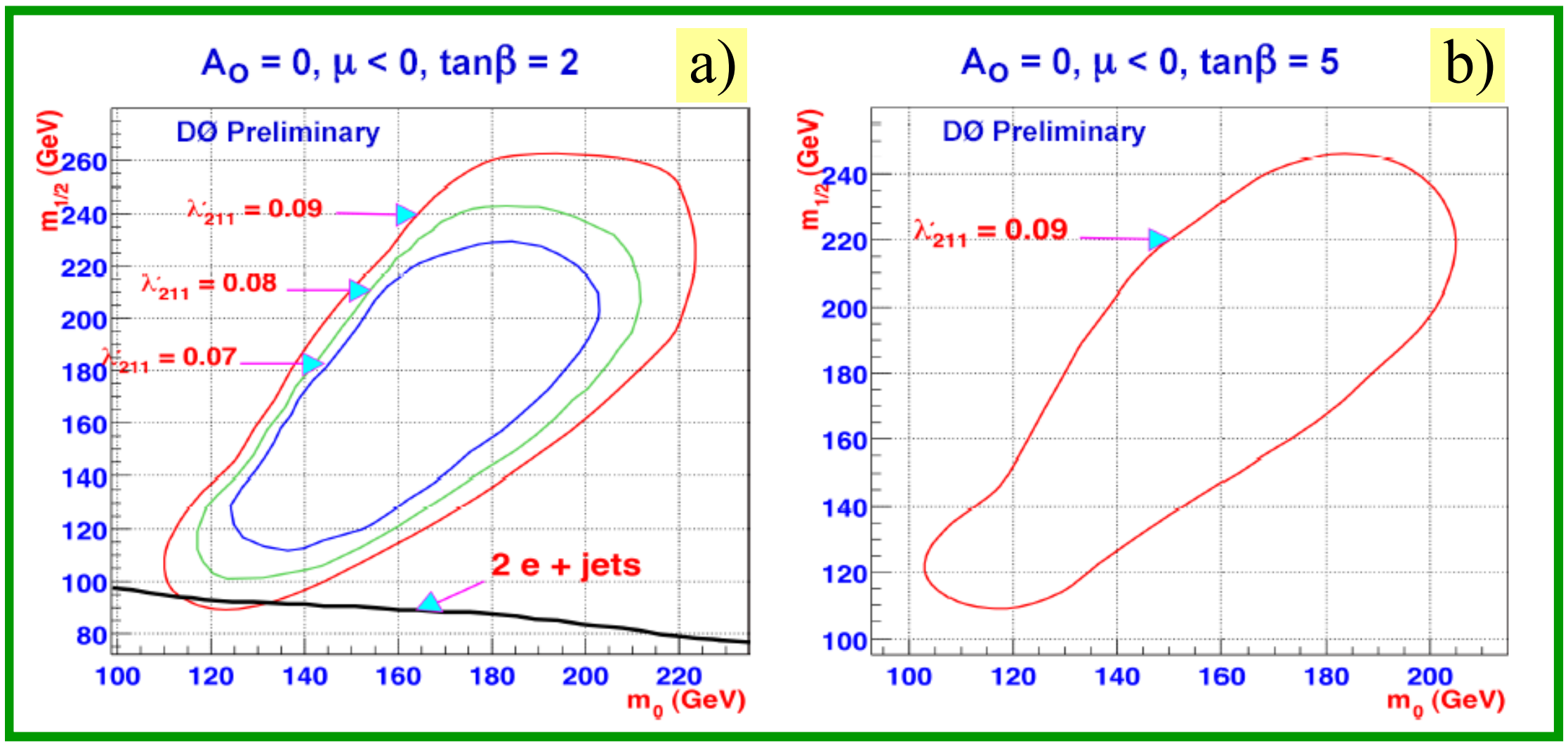

**FIG 5.** *Search for Resonant Slepton production in RPV mSUGRA. Exclusion curve as function of* $m_{1/2} \times m_0$ *for* $tg\beta = 2$ *and* $tg\beta = 6$.

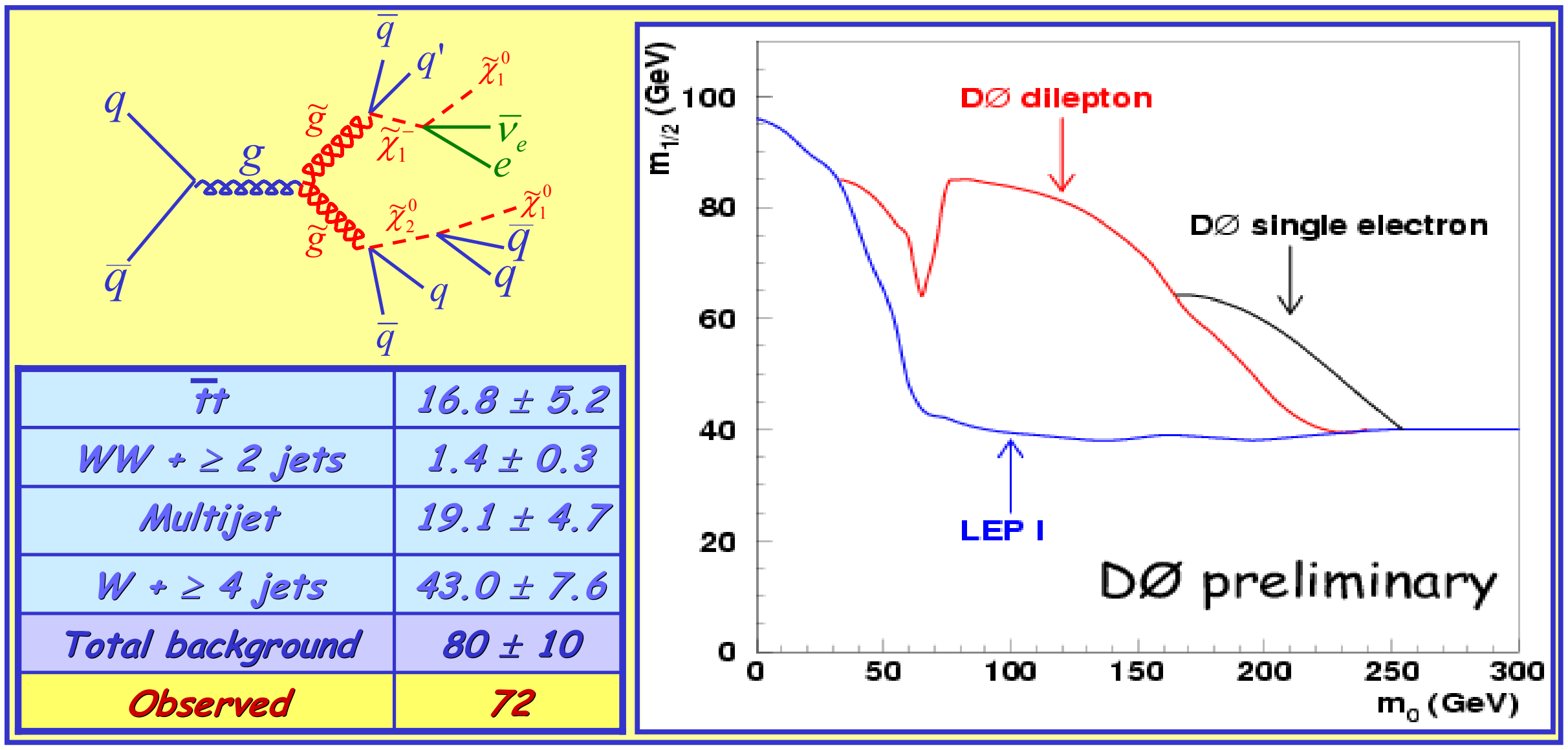

| | |
|---|---|
| $t\bar{t}$ | 16.8 ± 5.2 |
| WW + ≥ 2 jets | 1.4 ± 0.3 |
| Multijet | 19.1 ± 4.7 |
| W + ≥ 4 jets | 43.0 ± 7.6 |
| Total background | 80 ± 10 |
| Observed | 72 |

**FIG 6.** *Search for RPC mSUGRA in single electron channel. Exclusion curve is expressed as function of* $m_{1/2} \times m_0$.

## 5. Search for RPC mSUGRA in single electron channel

This process is sensitive to moderate $m_0$ region and complements di-lepton and $jets + \not{E}_T$ searches. The Feynman diagram from the process under investigation is shown in figure 6. The main backgrounds come from $t\bar{t}$, WW+jets, Multijet and W+jets events. The 95% C.L. limit curve, as obtained by D∅ Collaboration, is shown in Figure 6. The limits is represented in the usual $m_{1/2} \times m_0$ plane. Neural networks procedures have been also used in order to further optimize signal significance.

## 6. Search for Stop decay in 3-4 bodies

Stop decaying in 3 or 4 bodies have been searched recently by D∅ collaboration looking for final states containing an electron, a muon, missing transverse energy and jets in the final state. This is essentially a search for direct $\tilde{t}_1\bar{\tilde{t}}_1$ production where the stop decay in $\tilde{t}_1 \to b\ell\tilde{\nu} \to b\ell\nu\tilde{\chi}_1^0$. The main backgrounds of the analysis come from QCD multi-jet events, misidentified leptons, $WW \to \mu\nu e\nu$, $Z \to \tau\tau \to \mu e$, Drell-Yan and $t\bar{t}$ events. The search have been performed using 108 pb$^{-1}$ of data. The limits obtained are expressed as function of stop versus the neutralino mass and they are given in Figure 7.

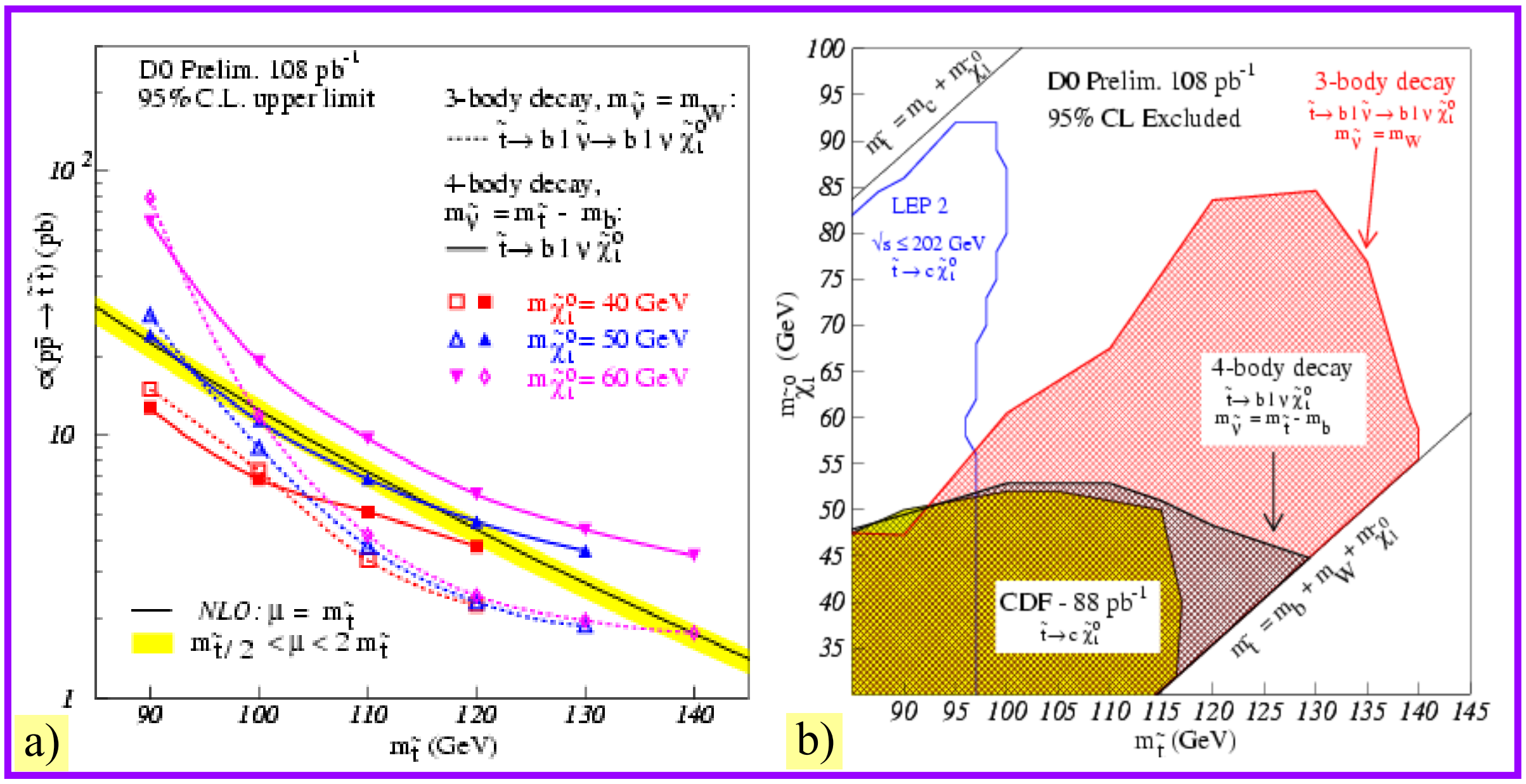

**FIG 7.** *Search for scalar top quark decaying in 3-4 bodies (D∅): 95% C.L. limit on the stop quark mass as function of the neutralino mass.*

## 7. Conclusions

Tevatron Experiments performed extensive searches for physics beyond the Standard Model using the data collected during the 1992-1996 Run I. Recent results on such searches in the framework of SUSGRA have been reported. No evidence for physics beyond the Standard Model have been found so far, then 95% C.L. limit have been set for the different scenarios described in the present paper.

## 8. Acknowledgments

I wish to thank the Organizers of the 10th International Conference on Supersymmetry and Unification of Fundamental Interactions for the excellent conference and their kind hospitality. The author wishes also to thank Teruki Kamon for his comments on the manuscript.

## 9. References

[1] H. Dreiner, **Pramana 51**, 123 (1998).
[2] W. Porod, D. Restrepo and J. W. Valle, hep-ph/0001033.
[3] R. Barate *et al.* [ALEPH Collaboration], **Eur. Phys. J. C 19**, 415 (2001).
[4] B. Abbott et al. [D∅ Collaboration], Phys.Rev.Lett. 83, 4476 (1999).
[5] B. Abbott et al. [D∅ Collaboration], FERMILAB-PUB-01-352-E, Nov 2001. 10pp.